\documentclass[wssquare]{ws-procs975x65}

\newcommand\neededonlyforarXiv[2]{#1}

\neededonlyforarXiv{
  \providecommand\eprint[1]{\href{http://arXiv.org/abs/#1}{{\tt [\url{arXiv:#1}]}}}
  \usepackage{hyperref}
  \hypersetup{
    colorlinks=true,
    citecolor=blue,
    linkcolor=magenta,
    filecolor=cyan,      
    urlcolor=cyan,
  }
}{
  \providecommand\eprint[1]{{\tt [arXiv:#1]}}
  \providecommand\href[2]{\url{#1}} 
}

\newcommand{\rmd}{{\rm d}}

\newcommand{\CA}{{\cal A}}

\newcommand{\CD}{{\cal D}}

\newcommand{\CI}{{\cal I}}

\newcommand{\CR}{{\cal R}}

\newcommand{\CQ}{{\cal Q}}
\newcommand{\CW}{{\cal W}}

\newcommand{\average}[1]{\left\langle #1 \right\rangle_\CD}
\newcommand{\initaverage}[1]{\left\langle #1 \right\rangle_{\CD_{\rm \bf i}}}

\newcommand{\initial}[1]{{{#1}_{\mathbf i}}}

\newcommand{\diffd}{\mathrm d}

\providecommand\cqg{Classical and Quantum Gravity}

\providecommand\cqg{CQG}

\providecommand\prd{Physical Review D}

\providecommand\grg{General Relativity and Gravitation}

\providecommand\apjl{Astrophys.J.Lett.}                 

\providecommand\aap{Astron.~Astroph.}
\providecommand\apj{Astrophys.~J.}                 

\begin{document}
\hypersetup{pdfpagelabels=false,breaklinks=false,linktocpage}

\markboth{Jan J. Ostrowski, Thomas Buchert \& Boudewijn F. Roukema}
         {RZA and averaging problem}

\title{On the relativistic mass function and averaging in cosmology}

\author{Jan J. Ostrowski$^{1,2,\dagger}$, Thomas Buchert$^{2}$ and Boudewijn F. Roukema$^{1,2}$}

\address{$^1$Toru\'n Centre for Astronomy, 
  Faculty of Physics, Astronomy and Informatics,
  Grudziadzka 5,
  Nicolaus Copernicus University, ul. Gagarina 11, 87-100 Toru\'n
Poland\\
$^\dagger$E-mail: ostrowski~at~astro.uni.torun.pl}

\address{$^2$Universit\'e de Lyon, Observatoire de Lyon, Centre de
  Recherche Astrophysique de Lyon,\\ CNRS UMR 5574: Universit\'e
  Lyon~1 and \'Ecole Normale Sup\'erieure de Lyon,\\ 9 avenue Charles
  Andr\'e, F--69230 Saint--Genis--Laval, France\footnote{BFR: during
    invited lectureship; JJO: during long-term visit.}}

\begin{abstract}
The general relativistic description of cosmological structure formation is an important challenge from both the theoretical and the numerical point of views.  In this paper we present a brief prescription for a general relativistic treatment of structure formation and a resulting mass function on galaxy cluster scales in a highly generic scenario. To obtain this we use an exact scalar averaging scheme together with the relativistic generalization of Zel'dovich's approximation (RZA) that serves as a closure condition for the averaged equations.    
\end{abstract}

\keywords{Relativistic cosmology; cosmological mass function.}


\bodymatter

\section{Introduction}

It is reasonable to assume that a proper analytic model should accompany if not precede any $N$-body simulation attempt, since it gives us a deeper understanding of the physics behind the process considered. That was the case with the classical Zel'dovich approximation (ZA) \cite{ZA} and the Press--Schechter (PS) mass function formula,\cite{PS74} predictions of which were confirmed afterwards, to a plausible degree, by Newtonian $N$-body simulations. In this line of thought a relativistic form of ZA as a subclass of the first-order Lagrangian perturbation theory \cite{Buchert89pancake} has been systematically translated to the relativistic stage,\cite{BuchRZA1} generalizing the pioneering proposal by Kasai.\cite{Kasai95II} In an ongoing work we concentrate on the generalization of the mass function in this relativistic framework.
We build on earlier work on the generalization of the Newtonian mass function \cite{BKS00,KBF01} that essentially introduces 
the complete initial data set, i.e. not only the overdensity but the three scalar invariants of the velocity gradient, to describe collapsing structures. This framework contains attempts to generalize the PS framework to a triaxial anisotropic collapse, since it is in addition inhomogeneous.

\section{Averaging in cosmology }
Given a flow-orthogonal, synchronous foliation of space-time (that restricts the matter model to irrotational dust), the averaged evolution of a general inhomogeneous and restmass-preserving spatial domain is subject to an effective form of Friedmann's differential equations:\cite{Buch00scalav,Buch01scalav}
\begin{eqnarray}
\label{effective1}
\left( \frac{{\dot a}_\CD}{a_\CD}\right)^2 \;=\; \frac{8\pi G \varrho^\CD_{\rm eff}}{3} + \frac{\Lambda}{3} - \frac{k_\CD}{a_\CD^2} \;\;; \nonumber\\
\left(\frac{{\ddot a}_\CD}{a_\CD}\right)\;=\; - \frac{4\pi G (\varrho^\CD_{\rm eff}+3p^\CD_{\rm eff})}{3} + \frac{\Lambda}{3} \;\;;\nonumber \\
{\dot\varrho}^\CD_{\rm eff} + 3 \left(\frac{{\dot a}_\CD}{a_\CD}\right) \left(\varrho^\CD_{\rm eff} + p^\CD_{\rm eff}\right)\;=\;0\;\;,
\end{eqnarray}
where $a_{\CD}$ is the domain dependent scale factor defined as the cubic root of the domain's volume, and where the sources are defined as $\varrho^\CD_{\rm eff} = \average{\varrho} + \varrho^\CD_{\Phi}$ for the actual matter source $\average{\varrho}$ and the extra backreaction density $\varrho^\CD_{\Phi}$. Notice that backreaction is a result of averaging also the geometrical side of Einstein's equations. Averaging leads to an effective pressure $p^\CD_{\rm eff} = p_{\Phi}^\CD$ (Note that the matter model is still dust and the chosen foliation
of space-time is unchanged). The new backreaction sources are defined in terms of the backreaction variables $\CQ_\CD$ and $\CW_\CD : = :\average{\CR} - \frac{6k_{\CD}}{a_\CD^2}$, where this latter is the deviation of the averaged scalar curvature $\langle \CR \rangle_{\CD}$ from the homogeneous curvature term. For the backreaction sources we have:
\begin{eqnarray}
\label{effective2}
\varrho_{\Phi}^{{\CD}} & := & -\frac{1}{16\pi G}{\CQ}_{{\CD}}-\frac{1}{16\pi G}\CW_\CD \;;\nonumber
\label{eq:equationofstate}\\
{p}_{\Phi}^{{\CD}} & := & -\frac{1}{16\pi G}{\CQ}_{{\CD}}+\frac{1}{48\pi G}\CW_\CD \;,
\end{eqnarray} 
allowing for an interpretation of the backreaction sources in terms of an effective scalar field.\cite{BuchLarAl06morph}
The kinematical backreaction term $\CQ_\CD$ is built from two extrinsic curvature invariants that are related to the kinematical invariants rate of expansion $\Theta$ and rate of shear $\sigma^2 : = \sigma^i_{\ j} \sigma^j_{\ i}$, with the shear tensor components $\sigma_{ij}$:
\begin{equation}
\label{backreaction}
{\CQ_{\CD}} =  2 \langle \rm{II} \rangle_{\CD}-\frac{2}{3}\langle \rm{I}\rangle^{2}_{\CD} \;\;;\;\; {\rm I} : = \Theta \;\;;\;\; {\rm II} : = \frac{1}{3} \Theta^2 - \sigma^2 \;\;. \nonumber
\end{equation}
Equations~(\ref{effective1}) and Friedmann's are the same up to the dependence on the averaging domain; they are strictly the same if we postulate that homogeneous sources $\varrho_h (t)$ and $p_h (t)$ describe the average evolution, as is conjectured in the standard model and proved to hold in Newtonian cosmology.\cite{BuchertEhlers97} In general relativity this no longer holds true due to the non-conservation of the averaged curvature.\cite{BuchCarf08}


\section{Relativistic Zel'dovich approximation and its average}
The effective equations~(\ref{effective1}) can be closed by providing a dynamical equation of state that relates the effective sources.
The relativistic Zel'dovich approximation (RZA) \cite{BuchRZA1} provides such a closure. It prescribes a perturbation ansatz for 
Cartan co-frames,
\begin{equation}
  \boldsymbol{\eta}^{\, a}=\boldsymbol{\eta}_{H}^{\;\; a}+a(t){\bf P}^{\, a}\;\;;\;\; a=1,2,3 \;\;,
\label{ansatz*}
\end{equation}
where $\boldsymbol{\eta}_{H}^{\;\; a}=\eta_{H\,
    i}^{\;\; a}\,\boldsymbol{\rmd}X^{i}:=a(t)\boldsymbol{\eta}_{H}^{\;\;
    a}(\initial t)\;,\;\eta_{H\, i}^{\;\; a}:=a(t)\delta^{\; a}_{\;\; i}$
describes the background deformation in the exact basis $\boldsymbol{\rmd}X^i$, $a(t)$ obeys the standard Friedmann equations, and the inhomogeneous
deformation one-form fields ${\bf P}^{\, a}(t,X^{k}) = P^a_{\;\,i}\,\boldsymbol{\rmd}X^i$ may be developed into a perturbation series.\cite{BuchRZA3}
In coordinate components and at first order, RZA has the form:
\begin{equation}
 ^{{\rm RZA}}\eta_{\ i}^{a}(X^{k}, t):=a(t)\left(\delta_{\ i}^{a}+\xi(t)\dot{P}_{\ i}^{a} (X^k , \initial{t})\right)\ ,
\label{eq:etaRZA}
\end{equation}
with $\xi (\initial{t}) = 0\;;\;\dot{\xi} (\initial{t}) = 1$.
The kinematical backreaction functional for this Lagrangian deformation field can be calculated:\cite{BuchRZA2}
\begin{eqnarray}
{}^{\rm RZA}{\cal Q}_\CD & = &  \frac{ \dot{\xi}^2 \left( \gamma_1 + \xi \gamma_2 +  \xi^2 \gamma_3 \right)}{\left(1 +  \xi \left\langle {\rm I}_{\rm \bf i} \right\rangle_{\cal I} + \xi^2 \left\langle {\rm II}_{\rm \bf i} \right\rangle_{\cal I} + \xi^3 \left\langle {\rm III}_{\rm \bf i} \right\rangle_{\cal I} \right)^2}\;.\nonumber
\end{eqnarray}
We here defined a formal (`initial') average, normalized by the initial volume,
\begin{equation}
\label{formalaverageB}
\left\langle \CA \right\rangle_\CI : =  \frac{1}{V_{\initial\CD}} \int_\CD \CA \sqrt{G} \,d^3 X \;;\;
V_{\initial\CD}  =  \int_{\initial\CD} \sqrt{G} \,d^3 X \;,
\end{equation}
where $G := \det(G_{i j})$ is taken on initial data for the metric coefficients, $G_{i j} (X^k) := g_{i j} (X^k, \initial{t})$;  the subscript $\mathrm i$ marks the initial data and the abbreviations stand for
\begin{eqnarray}
\gamma_1 = 2  \left\langle {\rm II}_{\rm \bf i} \right\rangle_{\cal I} - \frac{2}{3}  \left\langle {\rm I}_{\rm \bf i} \right\rangle_{\cal I}^2 = \CQ_{\initial\CD}\;; \nonumber \\
\gamma_2 = 6  \left\langle {\rm III}_{\rm \bf i} \right\rangle_{\cal I} - \frac{2}{3}  \left\langle {\rm I}_{\rm \bf i} \right\rangle_{\cal I}  \left\langle {\rm II}_{\rm \bf i} \right\rangle_{\cal I} \;;\nonumber \\
\gamma_3 = 2  \left\langle {\rm I}_{\rm \bf i} \right\rangle_{\cal I}  \left\langle {\rm III}_{\rm \bf i} \right\rangle_{\cal I} - \frac{2}{3}  \left\langle {\rm II}_{\rm \bf i} \right\rangle_{\cal I}^2 \;.
\end{eqnarray}
Note that, despite the approximation made, this functional is exact for special plane-symmetric inhomogeneities and for spatially flat LTB solutions.\cite{BuchRZA2} The corresponding functional in Newtonian cosmology \cite{BKS00} has the same form but contains the {\it{general}} plane- and spherically-symmetric solutions.

\section{Relativistic mass function}
Using RZA as a closure condition neglects pressure, velocity
dispersion and vorticity, which are most relevant at small scales,
as
 collapse processes accelerate. Generalizations of the matter
model are in progress.  
 In the following we will use the assumption
that all of the dark matter particles at $z=0$ are part of the dark
matter halos (following the Press--Schechter treatment), and that all
of the dark matter halos were formed from Gaussian distributions
in each of the initial invariants ($\initaverage{{\rm I}}$, $\initaverage{{\rm
    II}}$ and $\initaverage{{\rm III}}$)
with variances equal to those of the expectation values of the
variance of these invariants, respectively
(cf.\ App.C.2, \cite{BKS00}).  
We will also ignore the
cloud-in-cloud problem since it mostly affects the lower mass scale of
the mass function. 
Let us introduce the following notation: $n(z,M_i)$: the number of halos of mass $M_i$ at redshift $z$ per unit volume; $\hat{n}(M_i)$: the number of halos today; $F_i(z,M_i)$: the probability that objects of a given mass collapsed until the redshift $z$. 
$F_i(z,M_i)$ is calculated by a Monte Carlo procedure as explained in the corresponding Newtonian work.\cite{BKS00,KBF01} Since the probability of collapse is calculated independently for each mass scale, and because we assume that all of the mass at redshift $z=0$ is part of the collapsed objects, we need to normalise $F_i(0,M_i)$ so that (assuming $M_i = \rho_{\mathrm H}(t_{\mathrm{in}})\frac{4}{3}\pi (R_{\mathrm{in}})^3 $, where the superscript `in' stands for initial):
\begin{equation}
\int_{R_l}^{R_u} \alpha F_i(0,R) \diffd R = 1 \;,
\end{equation} 
where $R_l$ and $R_u$ correspond to lower (5 Mpc/h) and upper (80 Mpc/h) co-moving cut-offs, respectively, and $\alpha$ is the normalisation factor. 
This normalisation allows us to calculate the probability density of collapse for the given mass scale under the condition that all the mass in collapsed structures today adds up to the total mass of the domain containing these collapsed objects:
\begin{equation}
\label{norm}
\bar{F}_i(z,M_i) = \alpha F_i(z,M_i) \;.
\end{equation}
The number of collapsed objects in an arbitrary volume $V_{\mathrm H}$ is then given by: 
\begin{equation}
n(z,M_i) = \bar{F}_i(z,M_i)\frac{\rho_{\mathrm H} \;V_{\mathrm H}}{M_i} \;,
\end{equation}
where $\rho_{\mathrm H}$ is an average density (in our case the density of an EdS background Universe). Assuming that $M_i = \rho_{\mathrm H} (t_{\mathrm{in}})\frac{4}{3}\pi (R_{\mathrm{in}})^3$, we can rewrite the above equation:
\begin{equation}
\frac{n(z,M_i)}{V_{\mathrm H}}  = \bar{F}_i(z,M_i) \frac{\rho_{\mathrm H}}{\rho_{\mathrm H}(t_{\mathrm{in}})}\left(\frac{1}{\frac{4}{3}\pi (R_{\mathrm{in}})^3}\right) \,.
\end{equation}

\section{Results}
In this section we compare two cases of collapse models: spherical (no kinematical backreaction) and generic;
both cases start from Gaussian distributions in the initial invariants.
\begin{figure}[h]
\centering
\begin{minipage}{0.49\textwidth}
\includegraphics[width=\linewidth]{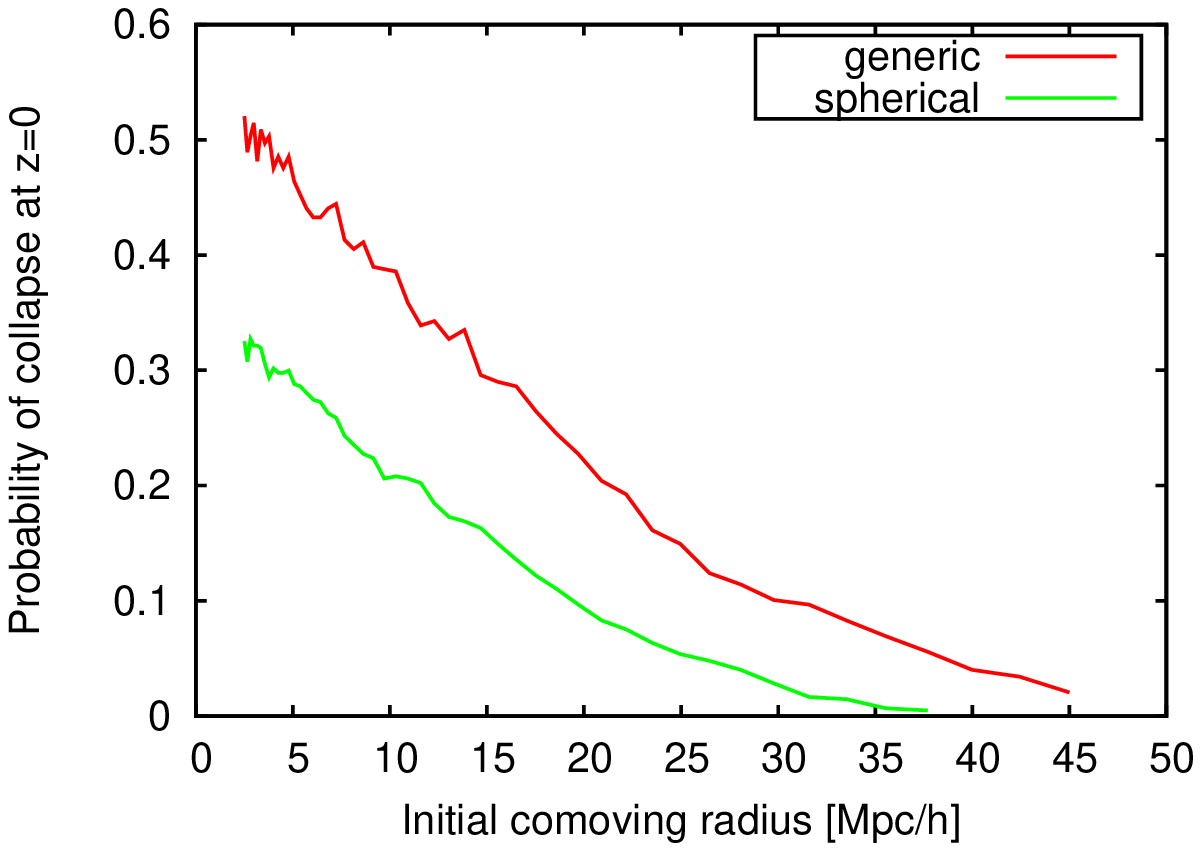}
\end{minipage}
\begin{minipage}{0.49\textwidth}
\centering
\includegraphics[width=\linewidth]{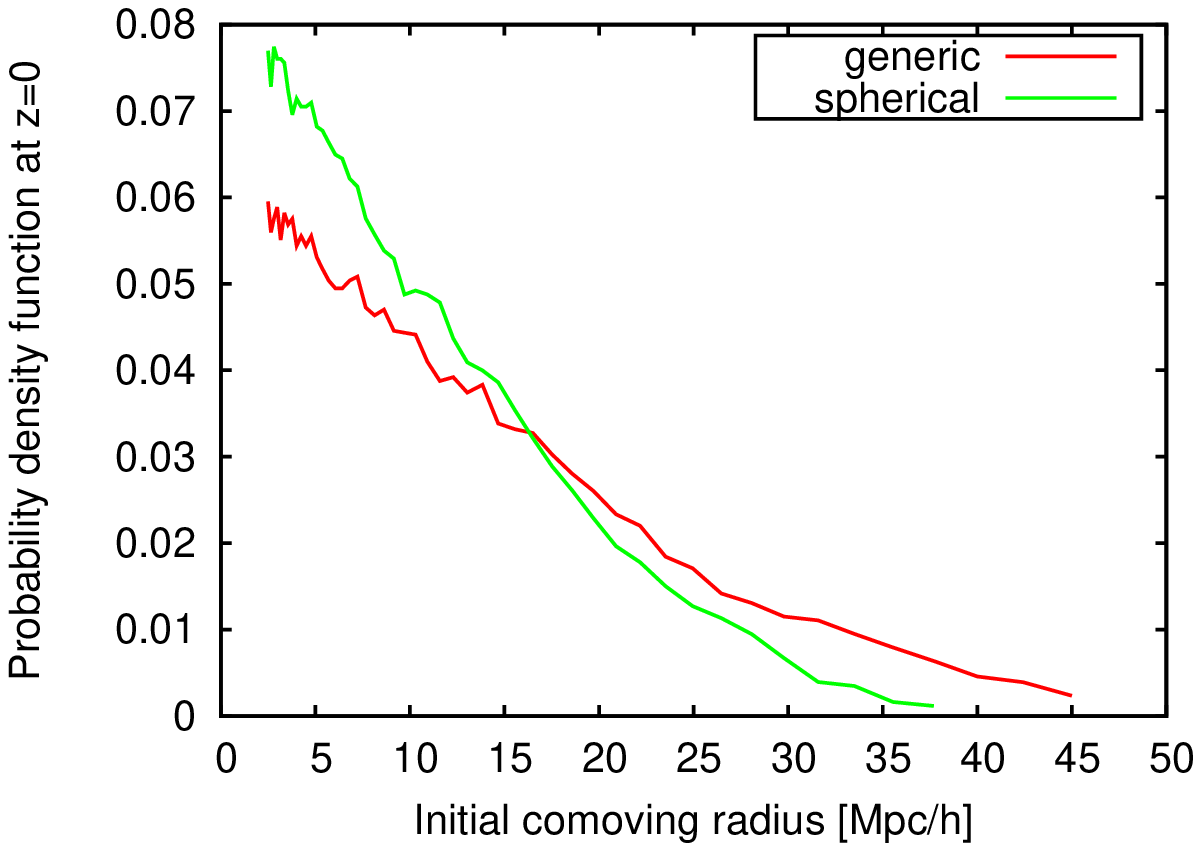}
\end{minipage}
\caption{Non-normalised probabilities and normalised (according to Eq.~(\ref{norm})) probability density function for the collapse of objects as a function of initial radius, at redshift $z=0$.}
\end{figure}
Figure~1 (left panel) shows that the individual probabilities of collapse are always higher for the generic case compared with the spherical model (this result has been also observed using different, non-spherical but less general approaches). The shear and the domain-dependent expansion rate accelerate the collapse, allowing bigger structures to form. Figure~1 (right panel) shows that making the assumption that all mass resides in collapsed objects at redshift $z=0$ changes the relation between these models---since bigger structures form in the generic case, less dark matter particles go into the low-mass end of the probability density function in comparison to the spherical case.  
\vspace{-10pt}
\begin{figure}[h]
\centering
\begin{minipage}{0.49\textwidth}
\includegraphics[width=\linewidth, height =.8\linewidth]{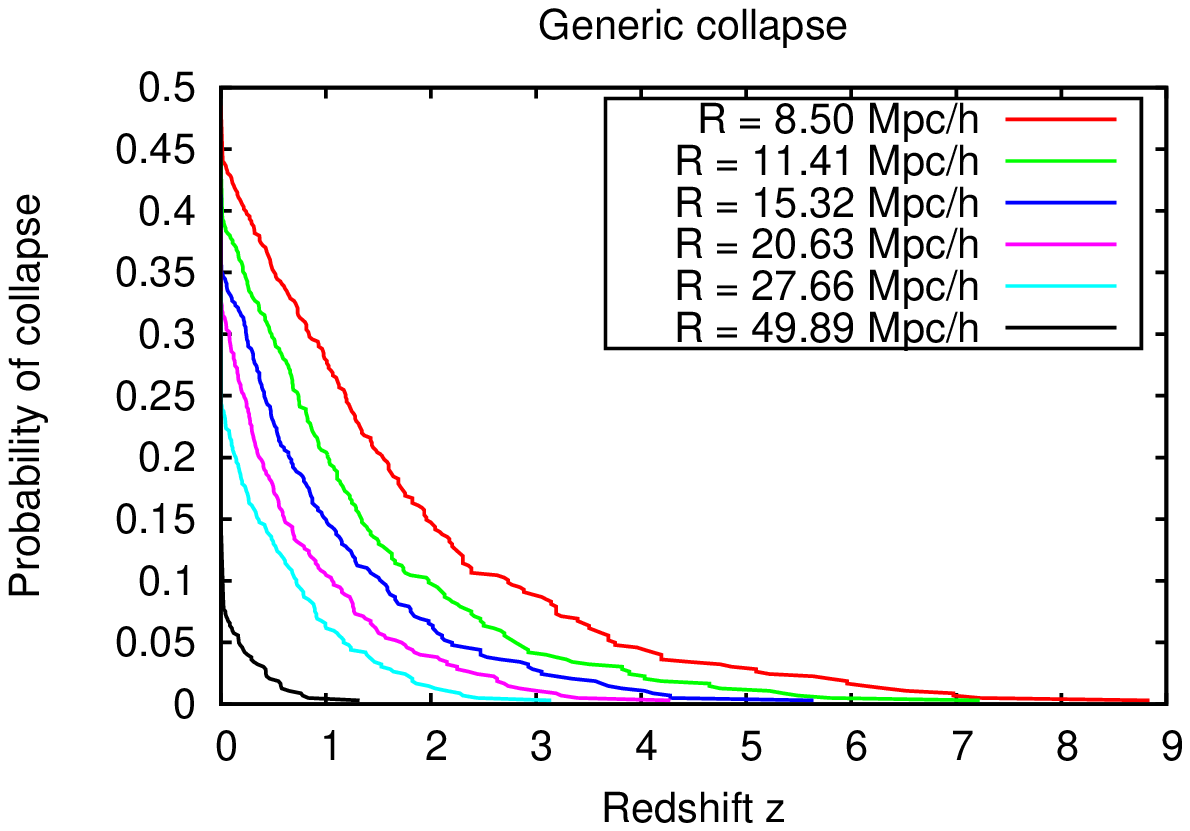}
\end{minipage}
\begin{minipage}{0.49\textwidth}
\centering
\includegraphics[width=\linewidth, height =.8\linewidth]{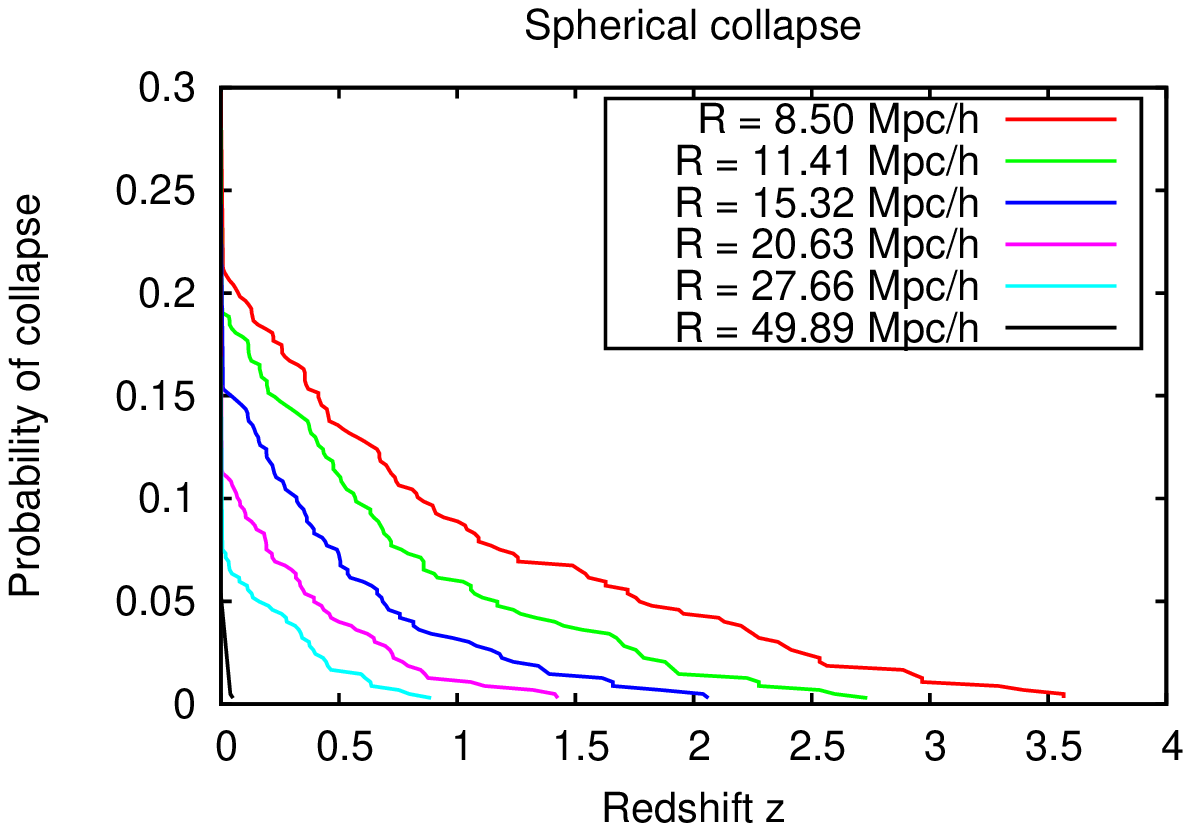}
\end{minipage}
\caption{Redshift-dependent probability of collapse for different mass scales.}
\end{figure}
\vspace{-20pt}
\section{Conclusions and Outlook}
Spherical collapse, being oversimplified, does not provide a plausible
analysis of structure formation compared with the generic situation
presented here. However, although individual probabilities of collapse
typically differ between the models by a factor of about two ($1.72$ at initial $R = 8$~Mpc/$h$ comoving),
when separately normalised to make all mass collapse by the present,
this ratio drops to about unity.
Taking into account the backreaction term results in higher abundances of collapsed objects at higher redshifts, and allows for a bigger bound structures to form (Fig.~2).  However, predictions concerning the low-mass end of the probability density function have to be taken with caution, since, as stated above, a better matter model is required to properly access the low mass regime.

In a work in preparation \cite{OstrowskiBuchR16} we also aim at
understanding the role of curvature in the distribution of collapsed
objects. We know from earlier considerations \cite{WiegBuch10} that
collapsed objects reside in positive-curvature environments. Since we
include backreaction in the generic model, we can quantify the
prediction that positive curvature energies add up to the effective matter
source, providing a scale-dependent abundance of a component that
would be interpreted as dark matter in the standard model.
By assuming purely baryonic matter content in the initial power 
spectrum, instead of the normal assumption that the
matter component is dominated by non-baryonic dark matter, 
the roles of matter content and curvature effects can be separated.

\section*{Acknowledgments}
We would like to thank Gilles Chabrier, Martin Kerscher, Fosca Al Roumi and Herbert Wagner for interesting discussions.
This work was conducted within the ``Lyon Institute of Origins'' under grant ANR--10--LABX--66. 
All authors acknowledge support from OPUS-7 grant 2014/13/B/ST9/00845 of the National Science Centre, Poland. 
Some of the calculations have been carried out in the context of grant 197 at the Pozna\'n Supercomputing and Networking Center (PCSS).


\begin{thebibliography}{10}

\bibitem{ZA}
Y.~B. {Zel'dovich}, {\em \aap} {\bf 5}, 84 (March 1970).

\bibitem{PS74}
W.~H. {Press} and P.~{Schechter}, {\em \apj} {\bf 187}, 425 (February 1974).

\bibitem{Buchert89pancake}
T.~{Buchert}, {\em \aap} {\bf 223}, 9 (October 1989).

\bibitem{BuchRZA1}
T.~{Buchert} and M.~{Ostermann}, {\em \prd} {\bf 86}, p. 023520 (July 2012),
  \eprint{1203.6263}.

\bibitem{Kasai95II}
M.~{Kasai}, {\em \prd} {\bf 52}, 5605 (November 1995).

\bibitem{BKS00}
T.~{Buchert}, M.~{Kerscher} and C.~{Sicka}, {\em \prd} {\bf 62}, p. 043525
  (August 2000), \eprint{astro-ph/9912347}.

\bibitem{KBF01}
M.~{Kerscher}, T.~{Buchert} and T.~{Futamase}, {\em \apjl} {\bf 558}, L79
  (September 2001), \eprint{astro-ph/0007284}.

\bibitem{Buch00scalav}
T.~{Buchert}, {\em \grg} {\bf 32}, 105 (January 2000), \eprint{gr-qc/9906015}.

\bibitem{Buch01scalav}
T.~{Buchert}, {\em \grg} {\bf 33}, 1381 (August 2001), \eprint{gr-qc/0102049}.

\bibitem{BuchLarAl06morph}
T.~{Buchert}, J.~{Larena} and J.-M. {Alimi}, {\em \cqg} {\bf 23}, 6379
  (November 2006), \eprint{gr-qc/0606020}.

\bibitem{BuchertEhlers97}
T.~{Buchert} and J.~{Ehlers}, {\em \aap} {\bf 320}, 1 (April 1997).

\bibitem{BuchCarf08}
T.~{Buchert} and M.~{Carfora}, {\em \cqg} {\bf 25}, 195001 (October 2008),
  \eprint{0803.1401}.

\bibitem{BuchRZA3}
A.~{Alles}, T.~{Buchert}, F.~{Al~Roumi} and A.~{Wiegand}, {\em \prd} {\bf 92},
  p. 023512 (July 2015), \eprint{1503.02566}.

\bibitem{BuchRZA2}
T.~{Buchert}, C.~{Nayet} and A.~{Wiegand}, {\em \prd} {\bf 87}, p. 123503 (June
  2013), \eprint{1303.6193}.

\bibitem{OstrowskiBuchR16}
J.~J. {Ostrowski}, T.~{Buchert} and B.~F. {Roukema}, {\em {in preparation}}
  (2016).

\bibitem{WiegBuch10}
A.~{Wiegand} and T.~{Buchert}, {\em \prd} {\bf 82}, 023523 (July 2010),
  \eprint{1002.3912}.

\end{thebibliography}
\end{document}